\documentstyle[prl,aps,twocolumn,graphicx]{revtex}

\newcommand{\eq}[1]{Eq.~(\ref{#1})}
\newcommand{\fig}[1]{Fig.~\ref{#1}}
\newcommand{\figs}[1]{Figs.~\ref{#1}}
\newcommand{\kappav}{\mbox{\boldmath\(\kappa\)}} % magnetic curv
\newcommand{\kappan}{{\kappa_{\mathrm{n}}}}    % normal curv
\newcommand{\kappag}{{\kappa_{\mathrm{g}}}}    % geodesic curv
\newcommand{\dotv}{\mbox{\boldmath\(\cdot\)}}  % dot product
\newcommand{\cross}{\mbox{\boldmath\(\times\)}}% cross product
\newcommand{\Bvec}{{\mathbf{B}}}               % magnetic field
\newcommand{\AGC}{{\cal{A}}}                   % ballooning term 1
\newcommand{\KGC}{{\cal{K}}}                   % ballooning term 2
\newcommand{\NGC}{{\cal{N}}}                   % ballooning term 3
\renewcommand{\d}{{\, \mathrm{d}}}             % differential
\newcommand{\grad}{\mbox{\boldmath\(\nabla\)}} % gradient
\newcommand{\irs}{{\cal{R}}}                   % int. residual shear
\newcommand{\J}{{\cal{J}}}                     % Jacobian
\newcommand{\smodel}{{\widehat{s}}}            % s-a shear
\newcommand{\alphamodel}{{\widehat{\alpha}}}   % s-a pressure
\newcommand{\vs}{\mbox{\it vs.\ }}             % verses

\begin{document}

\draft

\title{Anderson-Localized Ballooning Modes \\
  in General Toroidal Plasmas}
\author{
  P.\ Cuthbert and R.~L.\ Dewar\\ }
\address{Dept.\ of Theoretical Physics \& Plasma Research Lab., \\
  Research School of Physical Sciences and Engineering, \\
  The Australian National University, Canberra, Australia 0200}
\date{Preprint: \today}

\maketitle

\begin{abstract}

Ballooning instabilities are investigated in three-dim\-ens\-ional
magnetic toroidal plasma confinement systems with low global
magnetic shear.
The lack of any continuous symmetry in the plasma equilibrium
can lead to these modes being localized along the field lines
by a process similar to Anderson localization.
This produces a multibranched local eigenvalue dependence, where
each branch corresponds to a different unit cell of the extended
covering space in which the eigenfunction peak resides.
These phenomena are illustrated numerically for the
three-field-period heliac H-1, and contrasted with an axisymmetric
$s$-$\alpha$ tokamak model.
The localization allows a perturbative expansion about zero shear,
enabling the effects of shear to be investigated.

\pacs{PACS numbers: 52.35.P, 52.55.H, 71.23.A}

\end{abstract}

\narrowtext
%\twocolumn

Ballooning instabilities are pressure-driven ideal
magnetohydrodynamic (MHD) instabilities which limit the maximum
$\beta$ (plasma pressure/magnetic pressure) that can be obtained
in a plasma.  They are localized about regions where the field
lines are concave to the plasma, which are known as
{\em unfavourable} regions of curvature.  Another localizing
influence is the magnetic shear, which measures the rate at which
neighboring field lines at different minor radii separate
as they wind their way around the torus.  Large shear
helps stabilize these modes, thereby playing an important
role in the MHD stability.  In this paper however we consider the
effects of very small or zero shear, such as occurs in the heliac
class of stellarators or in the shear-reversal layers of an
advanced tokamak.

We begin by making the usual assumption that the magnetic field lines
map out nested flux surfaces, or {\em magnetic surfaces}.  These
are labeled using a normalized-toroidal-flux variable
$s$, which varies between zero at the center of the plasma and
unity at the plasma edge.
Within each surface the straight-field-line poloidal $\theta$
and toroidal $\zeta$ angle variables are defined such that
the field lines appear as straight lines in the $(\theta, \zeta)$
plane.  The magnetic field may then be written
$\Bvec = \grad \zeta \cross \grad \psi -
q \grad \theta \cross \grad \psi
\equiv \grad \alpha \cross \grad \psi$,
where the {\em field-line label} $\alpha \equiv \zeta - q \theta$.
Here, $2\pi \psi$ represents the {\em poloidal magnetic flux}, while
$q=q(s)$ is the {\em safety factor} (inverse of rotational transform),
which is equal to the average number of toroidal circuits traversed
by a field line per poloidal circuit traversed around the torus.

Ballooning modes can be characterized as having a long parallel and
short perpendicular wavelength with respect to the field lines.
By ordering the perpendicular wavelength to be small and expanding
to lowest order in an asymptotic series the local mode behavior can
be expressed by a one-dimensional equation along a field line
\cite{dewa83}.  Taking the plasma to be incompressible, the
ballooning equation may be written \cite{cuth98}
  \begin{equation}
    \left( \frac{\d}{\d \theta} \, \AGC \, \frac{\d}{\d \theta}
        - \KGC - \lambda \, \NGC \right) \xi = 0 \; ,
    \label{eq:ballooning}
  \end{equation}
where the eigenfunction $\xi$ is related to the mode displacement
while the eigenvalue $\lambda$ is equal to the mode growth rate
squared.  This represents the {\em local}
stability, local to a field line.  In forming {\em global} modes,
ray tracing must be performed in the three-dimensional $\lambda$
phase space to determine which of these local solutions also
satisfies the poloidal and toroidal symmetry requirements on the
phase of the mode \cite{coop96}.  This paper is concerned with
the functional dependence of $\lambda$ on its three arguments.

The ballooning coefficients are functions of local parameters and
$(\theta-\theta_k)$ ``secular'' terms, where the secular terms
represent the effects of magnetic shear
  \begin{eqnarray}
    \AGC &=& \frac{1}{\J |\grad \psi|^2}
      + \frac{|\grad \psi|^2}{\J B^2}
      [\irs + ( \partial_\psi q ) ( \theta - \theta_k ) ]^2 \; ,
    \label{eq:Adef} \\
    \KGC &=&
      -\frac{2 \J \partial_\psi p}{|\grad \psi|}
      \left\{ \kappan \frac{}{}
      \right. \nonumber \\
    & & \hspace{0.5cm} \left. 
      + \frac{|\grad \psi|^2}{B} \left[ \irs + ( \partial_\psi q )
      ( \theta - \theta_k ) \right] \kappag \right\} \; ,
    \label{eq:Kdef} \\
    \NGC &=& \J^2 \AGC \; ,
    \label{eq:Ndef}
  \end{eqnarray}
where $B$ represents the field strength, $p$ is the plasma
pressure, and the Jacobian is given by
$\J = ( \grad \psi \dotv \grad \theta \cross \grad \zeta {)}^{-1}$.
The local integrated shear is given by
$\irs + \theta \, \partial_\psi q =-\grad \alpha \dotv \grad \psi/
|\grad \psi|^2$, and the normal and geodesic
components of the magnetic curvature vector
$\kappav \equiv {\mathbf{e}_{||}} \dotv \grad {\mathbf{e}_{||}}$
(where ${\mathbf{e}_{||}} \equiv \Bvec / B$) are given by
$\kappan \equiv \kappav \dotv \grad s / |\grad s|$ and
$\kappag \equiv \kappav \dotv \grad s \cross \Bvec/|B\grad s|$,
respectively \cite{dewa84}.  The parameter $\theta_k$ is related
to the direction of the mode wave vector.

The coefficient $\AGC$ is positive definite, allowing the
ballooning equation to be transformed into the
Schr\"{o}dinger-like form
$\left[ \d^2 / \d \theta^2 + E - V \right] \AGC^{1/2} \xi = 0$,
where the ``potential''
  \begin{equation}
    V(\theta) = \frac{\KGC}{\AGC}
      - \frac{1}{4 \AGC^2} \left( \frac{\d \AGC}{\d \theta} \right)^2
      + \frac{1}{2 \AGC} \, \frac{\d^2 \AGC}{\d \theta^2} \; ,
  \end{equation}
and $E=-\lambda \J^2$ so that instability occurs whenever the
``energy'' is negative \cite{gree81}.

We first consider the case of a {\em two-dimensional} equilibrium,
which contains an ignorable third dimension.
Specifically we use the $s$-$\alpha$ tokamak model, which describes
an axisymmetric equilibrium analytically by assuming circular flux
surfaces and a large aspect ratio \cite{conn78}.  Being axisymmetric,
the toroidal angle is ignorable while the local parameters are
$2\pi$-periodic in the poloidal angle.
The potential may be written
  \begin{equation}
    V(\theta) = -\frac{\alphamodel \cos \theta}{1+h^2}
      + \frac{\left( h' \right)^2}
      {\left( 1+h^2 \right)^2 } \; ,
    \label{eq:salpha_V}
  \end{equation}
where $\smodel \equiv \partial (\ln q)/\partial (\ln r)$
is a measure of the global shear across the magnetic surface, 
$\alphamodel \equiv -2 R q^2 \partial_r p/B^2$
is a measure of the pressure gradient,
$h \equiv \smodel \, (\theta-\theta_k) - \alphamodel \sin \theta$
represents the integral of the local magnetic shear along the
field line, and $h' \equiv \d h / \d \theta$ is equal to the local
shear (we use $\smodel$ and $\alphamodel$ instead of the usual
labels $s$ and $\alpha$ in order to avoid confusion with the
normalized-toroidal-flux variable $s$ and the field-line label
$\alpha$).  Here, $r$ and $R$ represent the minor and major radii
of the torus, respectively.  The energy is given by
$E = -\lambda R q^2/B^2$.

%%%%%%%%%%%%%%%%%%%%%%%%
\begin{figure}
\includegraphics[bb=0.5in 4.75in 7.6in 9.85in, scale=0.45]
  {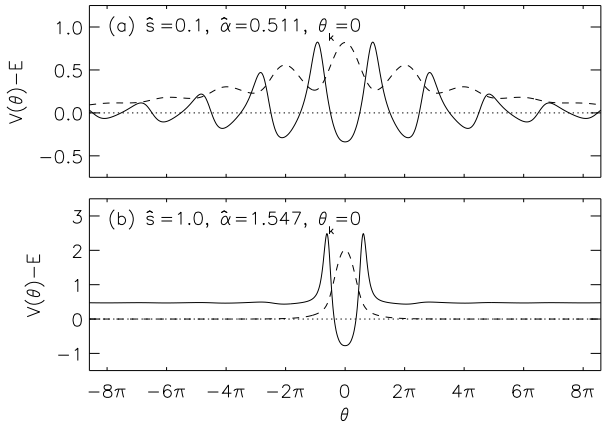}
\caption{Plots of $V(\theta)-E$ \vs $\theta$ (solid lines),
for the $s$-$\alpha$ model at $\theta_k=0$ and the most unstable
$\alphamodel$ for $\smodel=0.1$, 1.  Dashed lines represent the
wavefunctions $\AGC^{1/2} \xi$.}
\label{fig:s-alpha_pot_wave}
\end{figure}
%%%%%%%%%%%%%%%%%%%%%%%%

Two examples of $V(\theta)-E$ for the $s$-$\alpha$ model are shown
in \fig{fig:s-alpha_pot_wave}, along with the corresponding
``wavefunctions'' $\AGC^{1/2} \xi$, which are very similar to the
eigenfunctions $\xi$ for these cases.  These correspond to a
low-shear ($\smodel=0.1$) and a high-shear ($\smodel=1$) case.
The wavefunction (and eigenfunction) peaks occur at multiples of
$2\pi$ in $\theta$, where the normal curvature component
(given by $\kappan = -R^{-1} \cos \theta$)
is most unfavourable.  This behavior is modulated by
the secular terms, which are proportional to $\smodel$ and have
the effect of localizing the eigenfunction around
$\theta \approx \theta_k$.

The first term in \eq{eq:salpha_V} is alternatively stabilizing
and destabilizing as $\theta$ varies, being proportional to both
the normal curvature and the pressure
parameter.  This term is modulated by a function of the integrated
shear.  In the large-$|\theta|$ limit the influence of this shear
causes $V \rightarrow 0$, ensuring that all unstable solutions
(both the wavefunction and eigenfunction) are
exponentially localized.  Consequently, the $s$-$\alpha$ model
is stable to the more extended {\em interchange} modes.
The second term is purely stabilizing and represents the effects
of the local shear.  For large $|\theta|$ this term is
$O(\theta^{-4})$ compared with $O(\theta^{-2})$ for the first
term, so the effects of shear are actually more dominant in the
first term, with this term playing a significant role at more
moderate values of $\theta$.

%%%%%%%%%%%%%%%%%%%%%%%%
\begin{figure}
\includegraphics[bb=1in 4.75in 7.8in 9.85in, scale=0.5]
  {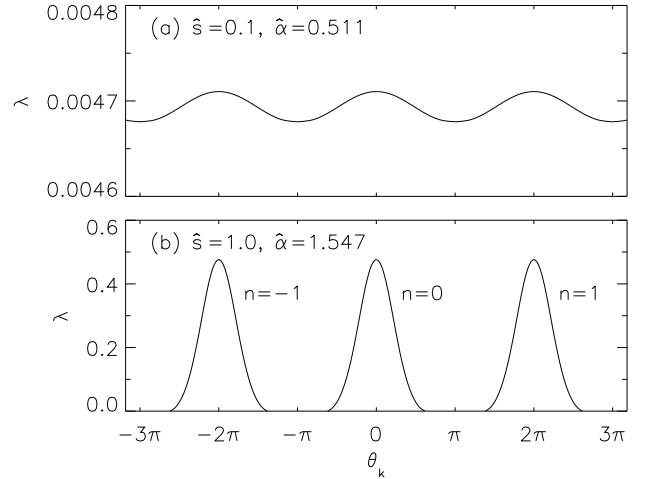}
\caption{Eigenvalues $\lambda$ \vs $\theta_k$, for the same
$\smodel$, $\alphamodel$ as \fig{fig:s-alpha_pot_wave}.
The poloidal branch labels $n$ are marked for the second case.}
\label{fig:s-a_evals_thkdepend}
\end{figure}
%%%%%%%%%%%%%%%%%%%%%%%%
    
In \fig{fig:s-a_evals_thkdepend} we investigate the
$\theta_k$-dependence of the local eigenvalues.  This is periodic
after a distance $2\pi$ in $\theta_k$, as expected from
the invariance of the ballooning equation under the poloidal
mapping operation, $P : \theta \mapsto \theta + 2\pi$,
$\theta_k \mapsto \theta_k + 2\pi$,
$\alpha \mapsto \alpha -2\pi q$.
An increase in $\theta_k$ by $2\pi$ is therefore associated with
an eigenfunction shift of $2\pi$ in the poloidal angle along the
field lines.  This enables us to define the
{\em poloidal branch label} $n$, representing the $2\pi$ interval
in $\theta$ where the eigenfunction is at a maximum.  Values of
this branch label are marked in \fig{fig:s-a_evals_thkdepend}(b),
where $n$ is defined such that the eigenfunction peak occurs
around $\theta \approx 2\pi n$.

The high-shear case gives well defined branches, in contrast to
the low-shear case where the branches merge so that there is
no stable interval of $\theta_k$.
The difference is due to the mode structure along
the field lines (see \fig{fig:s-alpha_pot_wave}).
As the shear is decreased the eigenfunctions become more extended
along the field lines and the difference between eigenfunctions
of neighboring branches is decreased.  In the
$\smodel \rightarrow 0$
limit all branches merge and the most unstable eigenfunction
becomes periodic after a distance $2\pi$ in $\theta$,
due to the vanishing of all secular terms.
The generalized solutions of the ballooning equation are then
{\em Bloch waves}, which can be written in the {\em Floquet} form
$\xi = \varphi_K(\theta) \exp({\mathrm{i}} K \theta)$, where $K$
represents the {\em Bloch wave vector} and $\varphi_{K}$
is a $2\pi$-periodic function in $\theta$.

In contrast to the above two-dimensional case, a general
three-dimensional equilibrium contains local parameters which
exhibit only {\em quasiperiodic} variation along the field lines,
since, when $q$ is irrational, a field line will cover a magnetic
surface ergodically.  We argue that this breaking of continuous
symmetry leads to the ballooning eigenfunctions being localized
along the field lines, even in the absence of magnetic shear.

We study an equilibrium which has the standard magnetic
configuration of the three-field-period H-1 heliac at the
Australian National University \cite{hamb90}.  The pressure
profile is similar to that which produces marginal stability
throughout the plasma \cite{coop94},
but increased by 20\% so that the plasma becomes highly unstable
with a $\beta$ of 1\% instead of the marginally stable value of
0.8\%.  The preconditioned {\tt VMEC} code \cite{hirs91} was used
to calculate the equilibrium and then a mapping code was used to
transform it into straight-field-line Boozer coordinates.
In doing this the poloidal and toroidal angles
were defined such that $(\theta,\zeta)=(0,0)$ corresponds to a
symmetry point on the outer side of the stellarator about which
stellarator symmetry holds \cite{dewa97c}.  This point occurs
deep within a region of unfavourable curvature.

%%%%%%%%%%%%%%%%%%%%%%%%
\begin{figure}
\includegraphics[bb=1.25in 5in 7.75in 9.67in, scale=0.5]
  {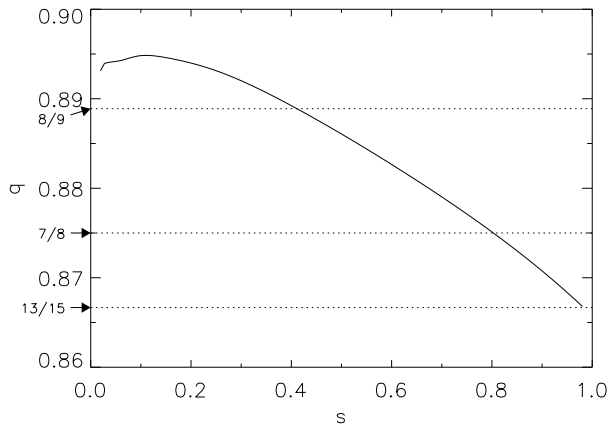}
\caption{Safety factor $q$ \vs surface label $s$, for the H-1
configuration studied.}
\label{fig:qplot_h1c}
\end{figure}
%%%%%%%%%%%%%%%%%%%%

A plot of the safety factor variation is shown as
\fig{fig:qplot_h1c}.  The shear is small throughout the plasma
volume with $q$ varying by less than 5\%.  The shear
parameter of the $s$-$\alpha$ model is approximately given by
$\smodel \approx 2 s \partial_s q / q$.
Using this expression, the maximum absolute value of $\smodel$
for this configuration is only around 0.12, and occurs at the
plasma boundary.

%%%%%%%%%%%%%%%%%%%%
\begin{figure}
\includegraphics[bb=1.25in 5.17in 7.75in 9.7in, scale=0.5]
  {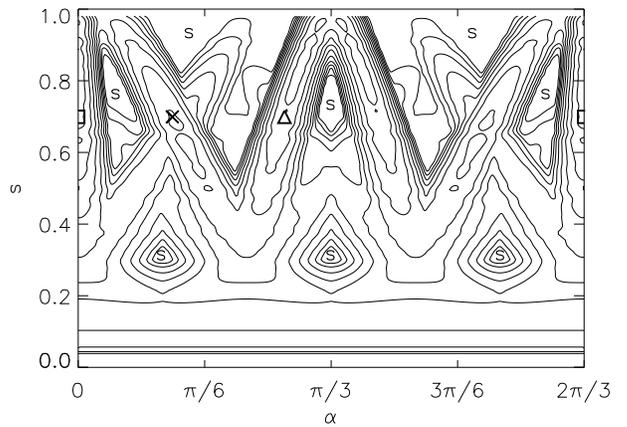}
\caption{Contours of local eigenvalues for
the H-1 configuration at $\theta_k=0$.  Areas labeled by `s'
represent stable regions.  Some corresponding eigenfunctions
are shown in \fig{fig:efns_h1c_is88}.}
\label{fig:contour_evals}
\end{figure}
%%%%%%%%%%%%%%%%%%%%

%%%%%%%%%%%%%%%%%%%%%%%%
\begin{figure}
\includegraphics[bb=1in 4.75in 7.5in 9.17in, scale=0.5]
  {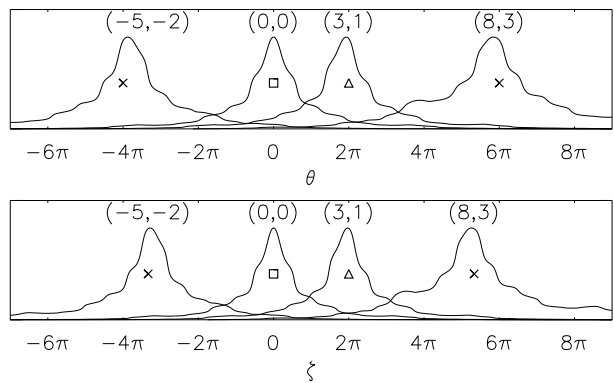}
\caption{Eigenfunctions $\xi$ \vs poloidal $\theta$ and
toroidal $\zeta$ angles, at the points marked by `$\cross$'
(two eigenfunctions), `$\Box$' and `$\triangle$' in
\fig{fig:contour_evals}.  The branch labels $(m,n)$ are marked.}
\label{fig:efns_h1c_is88}
\end{figure}
%%%%%%%%%%%%%%%%%%%%

A representation of the phase-space dependence of the (most
unstable) local eigenvalue for this configuration with
$\theta_k=0$ is shown as \fig{fig:contour_evals}, with
some corresponding eigenfunctions in \fig{fig:efns_h1c_is88}.
Despite the fact that the shear is small
($\smodel \approx -0.06$ at $s=0.70$),
the eigenfunctions are well localized
along the field lines and are mostly confined to a single
$2\pi \times 2\pi/M$ unit cell of the $(\theta, \zeta)$
covering space, where $M$ represents the number of
identical field periods in the stellarator ($M=3$ for H-1).
We use this to define the {\em poloidal} and
{\em toroidal branch labels}, which identify the unit cell
of the covering space where the eigenfunction has a maximum.
Each of the cases shown contains well defined branch labels,
due to the strong localization of the eigenfunctions.
This is in sharp contrast to the more extended eigenfunctions
of the $s$-$\alpha$ model at a similar value of $|\smodel|$
(see \fig{fig:s-alpha_pot_wave}(a)),
suggesting that something other than shear is responsible for
the localization.
This is confirmed by setting $\partial_\psi q \equiv 0$ in
\eq{eq:ballooning} to eliminate the secular terms altogether.
Once again, we find the eigenfunctions to be well localised,
and indeed to be almost identical to their counterparts in
\fig{fig:efns_h1c_is88}.

%%%%%%%%%%%%%%%%%%%%%%%%
\begin{figure}
\vspace{5mm}
\includegraphics[bb=1in 4.75in 7.5in 9.17in, scale=0.5]
  {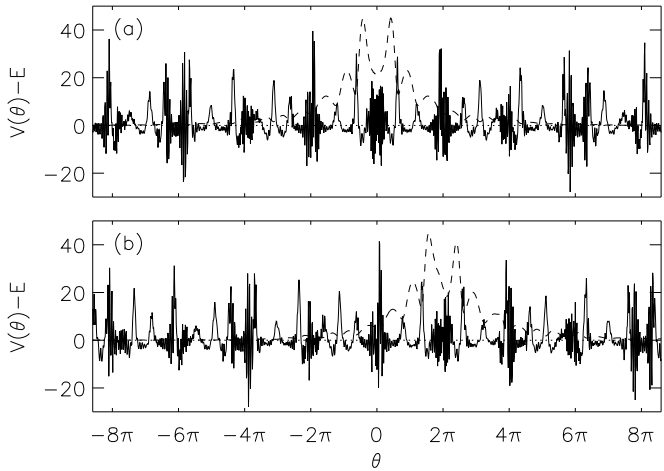}
\caption{Same as \fig{fig:s-alpha_pot_wave}, except that this
plot corresponds to the (a) `$\Box$' and (b) `$\triangle$'
eigenfunctions of \fig{fig:efns_h1c_is88}.  The small-scale
fluctuations in $V(\theta)$ are a consequence of ripple
effects from the 36 toroidal field coils.}
\label{fig:plot_sch3_efn_h1c}
\end{figure}
%%%%%%%%%%%%%%%%%%%%

This localization can be attributed to
{\em Anderson localization}, the process by which electron
wavefunctions become localized in space due to the presence of
impurities in an otherwise perfect crystal \cite{ande58}.
In this case however it is the quasiperiodicity of local
parameters, due to the incommensurate periods of the toroidal
and poloidal variations on a field line when $q$ is
irrational, which cause the localization \cite{sarn82,chul89}.
We plot two examples of the ``potential'' in
\fig{fig:plot_sch3_efn_h1c}, which is quasiperiodic in the
zero-shear limit.  From this plot it is not immediately clear
where the wavefunction $\AGC^{1/2} \xi$ will reside, since
potential ``wells'' exist at many places along the field lines.

The wave packet location and the resulting eigenvalue dependence
can be understood in terms of the normal curvature.  First
consider the $(m,n) = (0,0)$ branch, labeled by `$\Box$' in
\figs{fig:contour_evals} and \ref{fig:efns_h1c_is88}.  This
eigenfunction contains a maximum around the
$(\theta, \zeta) \approx (0,0)$ region of unfavourable curvature,
corresponding to $\alpha \approx 0$ in \fig{fig:contour_evals},
independent of $s$.  The $(m,n) = (3,1)$ branch on the other hand
appears to be most unstable around the
$\alpha \approx 2 \pi (m / M - n q ) \approx 0.76$ field line
when $s=0.70$, this being the field line that
passes through the same $(\theta, \zeta) \approx (0,0)$
unfavourable region, but only after traversing the torus once in
both the poloidal and toroidal directions.  Similarly, other
branches contain eigenfunctions which peak around the same region
of unfavourable curvature, but only after the field line has
undergone $m/M$ toroidal and $n$ poloidal transits around the torus.
The most-unstable field line of each branch with $n \ne 0$ will
therefore be a function of $q(s)$, resulting in a complex
``multibranched'' eigenvalue structure, such as that shown in
\fig{fig:contour_evals}.

While this behavior is qualitatively correct, in practise we
find a small discrepancy between the predicted and observed
positions of the maximum eigenvalue of each branch \cite{dewa99}.
For example, the $(3,1)$ branch is actually most unstable at
$\alpha \approx 0.85$ for $s=0.70$, in comparison with the
predicted value of $\alpha \approx 0.76$.  This ``shifting''
effect can be investigated using a two-dimensional perturbative
expansion of the ballooning equation, where we first solve along
the most-unstable field line $\alpha_0$ with the global shear
set to zero, and we then treat finite $(\alpha-\alpha_0)$,
$\partial_\psi q$ as a perturbation.

Provided $q$ is sufficiently {\em irrational}, the zeroth-order
eigenfunction estimate will be square integrable and will be
mostly confined to a single unit cell of the covering space.
We perturb this solution by including terms in both
$\partial_\psi q$ and $(\alpha-\alpha_0)$, which are assumed to be
of the same order, and we consider all branches by including the
branch labels explicitly.  The final eigenvalue estimate to second
order can then be written in the compact form
  \begin{eqnarray}
    \lambda & \approx &
      \lambda_0 + \epsilon \lambda_\epsilon
      + \epsilon^2 \lambda_{\epsilon^2}
      + \epsilon^2 \left( \theta_k - 2\pi n \right)^2
      \left( \lambda_{\epsilon^2 \theta_k^2} -
      \frac{\lambda_{\alpha \epsilon \theta_k}^2}
        {4 \lambda_{\alpha^2}} \right)
      \nonumber \\
    & & \hspace{-0.4cm} + \lambda_{\alpha^2} \left[ \alpha
      + 2\pi \left( n q - \frac{m}{M} \right)
      + \epsilon \left( \theta_k - 2\pi n \right)
      \frac{\lambda_{\alpha \epsilon \theta_k}}
        {2 \lambda_{\alpha^2}} \right]^2 ,
    \label{eq:lambda_expansion_n}
  \end{eqnarray}
where $\epsilon \equiv \partial_\psi q$, the $\lambda_x$
coefficients are functions of the magnetic surface only, and for
this H-1 configuration $\alpha_0=0$ so all functions are of
definite parity.  Each of the above terms has a simple physical
interpretation.  The first
represents the unperturbed estimate, which is corrected by
the following two terms which act on all branches equally.  These
are stabilizing and represent the effects of shear on the $(0,0)$
branch.  The fourth term is also stabilizing and represents the
effects of shear as a field line is followed through one or more
complete poloidal rotations.  Finally there is the last term,
which contains the field-line eigenvalue dependence and the
eigenvalue peak ``shift'' term.  This shift can be investigated
by ordering the eigenfunction to be well localized along the
field line, so that all odd-parity terms vanish at leading order.
To lowest order we obtain
$\irs + ( \partial_\psi q ) ( \theta - \theta_k ) \approx 0$
at the mode peak, showing that the most unstable part of the
branch occurs where the integral of the magnetic shear
approximately vanishes at the mode peak.

We have implemented this expansion numerically and have found
that in most cases the expansion provides good agreement with
eigenvalues calculated directly, except near the plasma edge and
around the $q=8/9$ rational surface (containing $6\pi$-periodic
local coefficients).  In particular, the predicted ``shifts'' in
the position of the eigenvalue maximum are in approximate agreement
with those observed directly and those obtained by assuming that
$\irs + ( \partial_\psi q ) ( \theta - \theta_k ) \approx 0$ at the
mode peak, supporting our interpretation of the mechanism behind
this term (details of these results will be reported elsewhere).

The numerical calculations were performed on the Australian
National University Supercomputer Facility's Fujitsu VPP300
vector processor.
We wish to thank Dr.\ Henry Gardner for the H-1 heliac {\tt VMEC}
input files and Dr.\ S.~P.\ Hirshman for the use of the {\tt VMEC}
code.


\begin{references}

\bibitem{dewa83}
R.~L.\ Dewar and A.~H.\ Glasser,
Phys.\ Fluids {\bf 26}, 3038 (1983).

\bibitem{cuth98}
P.\ Cuthbert, J.~L.~V.\ Lewandowski, H.~J.\ Gardner, M.\ Persson,
D.~B.\ Singleton, R.~L.\ Dewar, N.\ Nakajima, and W.~A.\ Cooper,
Phys.\ Plasmas {\bf 5}, 2921 (1998).

\bibitem{coop96}
W.~A.\ Cooper, D.~B.\ Singleton, and R.~L.\ Dewar,
Phys.\ Plasmas {\bf 3}, 275 (1996);
{\bf 3}, 3520(E) (1996).

\bibitem{dewa84}
R.~L.\ Dewar, D.~A.\ Monticello, and W.~N.~-C.\ Sy,
Phys.\ Fluids {\bf 27}, 1723 (1984).

\bibitem{gree81}
J.~M.\ Greene and M.~S.\ Chance,
Nucl.\ Fusion {\bf 21}, 453 (1981).

\bibitem{conn78}
J.~W.\ Connor, R.~J.\ Hastie, and J.~B.\ Taylor,
Phys.\ Rev.\ Lett.\ {\bf 40}, 396 (1978).

\bibitem{hamb90}
S.~M.\ Hamberger, B.~D.\ Blackwell, L.~E.\ Sharp, and D.~B.\ Shenton,
Fusion Technol.\ {\bf 17}, 123 (1990).

\bibitem{coop94}
W.~A.\ Cooper and H.~J.\ Gardner,
Nucl.\ Fusion {\bf 34}, 729 (1994).

\bibitem{hirs91}
S.~P.\ Hirshman and O.\ Betancourt,
J.\ Comput.\ Phys. {\bf 96}, 99 (1991).

\bibitem{dewa97c}
R.~L.\ Dewar and S.~R.\ Hudson,
Physica D {\bf 112}, 275 (1997).

\bibitem{ande58}
P.~W.\ Anderson,
Phys.\ Rev.\ {\bf 109}, 1492 (1958).

\bibitem{sarn82}
P.\ Sarnak,
Commun.\ Math.\ Phys.\ {\bf 84}, 377 (1982).

\bibitem{chul89}
V.~A.\ Chulaevsky and Ya.~G.~Sinai,
Commun.\ Math.\ Phys.\ {\bf 125}, 91 (1989).

\bibitem{dewa99}
R.~L.\ Dewar and P.\ Cuthbert,
Chinese Phys.\ Lett. (to be published).

\end{references}
\end{document}